\documentstyle[12pt]{article}
\def\nsp{\noindent}
\def\eea{\end{eqnarray}}
\def\bea{\begin{eqnarray}}
\def\eeas{\end{eqnarray*}}
\def\beas{\begin{eqnarray*}}
\def\ee{\end{equation}}
\def\be{\begin{equation}}

\def\fr{\frac}

\def\iso{\vec\tau}

\def\dag{^\dagger}

\def\fpi2{\mbox{F$_\pi$}^2}

\def\mpi2{{m_\pi}^2}
\def\mk{m_K}
\def\mk2{{m_K}^2}
\def\fkk{\mbox{F$_K$}}
\def\fk2{\mbox{F$_K$}^2}
\def\df{F^\prime}

\def\piv{\hat\pi}
\def\r{\vec r}

\def\Fpis{{F_\pi^2}}

\renewcommand{\thefootnote}{\fnsymbol{footnote}}
\begin{document}
\begin{titlepage}

\begin{center}

\vspace*{2.0cm}
\hfill TAN-FNT-98-01

\hfill LA PLATA TH-98-01
\vspace*{2.0cm}

{\large\bf STRANGE MULTISKYRMIONS\footnote{Work supported in
part by a grant of the Fundaci\'on Antorchas, Argentina.}}
\vskip 1.5cm

{Martin SCHVELLINGER$^a$ and
Norberto N. SCOCCOLA$^b$\footnote[2]{Fellow
of the CONICET, Argentina.}}
\vskip .2cm
{\it
$^a$ Physics Department, University of La Plata, C.C. 67, (1900)
La Plata, Argentina. \\
$^b$ Physics Department, Comisi\'on Nacional de Energ\'{\i}a At\'omica,
	  Av.Libertador 8250, (1429) Buenos Aires, Argentina.\\}

\vskip 2.cm
January 1998
\vskip 2.cm
{\bf ABSTRACT}\\
\begin{quotation}
Using a recently proposed approximation for multiskyrmion fields
based on rational maps we study the masses and baryonic radii
of some strange multibaryons within the bound state soliton model.
We find the tetralambda binding to be stronger than previously expected.
In addition, the model predicts the existence of a ``heptalambda" which
is stable against strong decays.
\end{quotation}
\end{center}

\vskip 1.cm


\end{titlepage}

\renewcommand{\thefootnote}{\arabic{footnote}}
\setcounter{footnote}{0}

\vskip 1cm

In recent years there has been a continuous effort in finding
minimum energy skyrmion solutions for increasing winding number $B$.
A decade ago, the so-called $B=2$ torus configuration was found\cite{KS87}.
A few years later the lowest $B=3$ and $B=4$
energy solution were identified\cite{BTC90}. Finally, very
recently some $B=5$ to 9 skyrmion configurations which are believed
to represent global minima were discovered\cite{BS97}. 
All these solutions
could be obtained only after a sophisticated and demanding numerical
work. The need to further investigate the physics behind these
solutions has lead people to find good and algebraically simple
approximations to them. For example, in Ref.\cite{KKSO89} a variational
approximation to the B=2 torus was introduced. Even more important
for our purposes, less than a year ago Houghton, Manton and 
Sutcliffe \cite{HMS97}
have exploited the similarities between the BPS monopoles and skyrmions
to propose some ans\"atze based on rational maps.
They have shown that
such configurations approximate very well the numerically found lowest
energy solutions with $B \leq 9$ and conjectured they could also
describe that with $B=17$. The introduction of these
approximations allows for the study of the properties of
multibaryons with strangeness. These exotics have received
quite a lot attention since Jaffe first speculated on the possible
existence of a stable H-dibaryon twenty years ago\cite{Jaf77}.
Strange dibaryons have been extensively studied in different models
including the Skyrme model in its various $SU(3)$ extensions
\cite{BBLRS84,KSS92,KM88,TSW94}. The tetralambda has been
also considered using the $SO(3)$ ansatz for the skyrmion
field\cite{IKS88}. In addition, experimental searches of these
objects are in progress, particularly in heavy ion collisions
at Brookhaven and CERN
(see i.e. Refs.\cite{E864} and references therein).
The purpose of the present work is to extend previous investigations
by considering strange multiskyrmions with $B=3$ to
$9$ and $17$ in the bound state
approach to the $SU(3)$ Skyrme model\cite{CK85}. In this approach strange
multibaryons can be obtained by binding kaons to a multisoliton background.
To describe
such multisolitonic field we will use the rational map ans\"azte.

We start with the effective action for the simple Skyrme model with an
appropriate symmetry breaking term, expressed in terms of the
$SU(3)$--valued chiral field $U(x)$ as
\bea
\Gamma = \int d^4x \ \left\{
{\Fpis \over{16}} Tr\left[ \partial_\mu U \partial^\mu U^\dagger \right] +
{1\over{32 e^2}} Tr\left[ [U^\dagger \partial_\mu U ,
U^\dagger \partial_\nu U ]^2 \right] \right\} + \Gamma_{WZ} +
\Gamma_{SB} \ ,  \label{action}
\eea
\nsp
where $F_\pi$ is the pion decay constant (~$= 186 \ MeV$ empirically)
and $e$ is the so--called Skyrme parameter.  In Eq.(\ref{action}),
the symmetry breaking term $\Gamma_{SB}$ accounts for the different
masses and decay constants of the pion and kaon fields while
$\Gamma_{WZ}$ is the usual Wess--Zumino action. Their explicit
forms can be found in i.e. Ref.\cite{TSW94}.

We proceed by introducing the Callan--Klebanov (CK) ansatz for the
chiral field
\be
\label{ansatz}
U=\sqrt{U_\pi}U_K\sqrt{U_\pi} \ .
\ee
In this ansatz, $U_K$ is the field that carries
the strangeness. Its form is
\bea
U_K \ = \ \exp \left[ i\fr{2\sqrt2}{{\fkk}} \left( \begin{array}{cc}
							0 & K \\
							K\dag & 0
						   \end{array}
					   \right) \right] \ ,
\eea
where $K$ is the usual kaon isodoublet
$
K \ = \ \left( \begin{array}{c}
		   K^+ \\
		   K^0
		\end{array}
			   \right) \ .
$

     The other component, $U_\pi$, is the soliton background field. It
is a direct extension to $SU(3)$ of the $SU(2)$ field $u_\pi$, i.e.,
\bea
U_\pi \ = \ \left ( \begin{array}{cc}
		       u_\pi & 0 \\
		       0 & 1
		    \end{array}
			       \right ) \ .
\eea

Replacing the ansatz Eq.(\ref{ansatz}) in the effective action
Eq.(\ref{action}) and expanding up to second order in the kaon
fields we obtain the lagrangian density for the kaon--soliton system.
In the spirit of the bound state approach this coupled system
is solved by finding first the soliton background configuration. For 
this purpose we introduce the rational map ans\"atze\cite{HMS97}
\be
u_\pi = \exp\left[ i \vec \tau \cdot \hat \pi_n \ F \right] \ ,
\label{mansatz}
\ee
with
\be
\hat \pi_n = {1\over{1+|R_n|^2}} \left( 2 \ \Re(R_n) \ \hat \imath +
				      2 \ \Im(R_n) \ \hat \jmath +
				     ( 1 - |R_n|^2 ) \ \hat k \right) \ ,
\label{pians}
\ee
where we have assumed that $F = F(r)$ and $R_n = R_n(z)$ is
the rational map corresponding to winding number $B=n$.
Here, $r$ is the usual spherical radial coordinate whereas the
complex variable $z$ is related to the other two spherical
coordinates $(\theta,\phi)$ via stereographic
projection, namely $z = \tan(\theta/2) \exp(i \phi)$.

In order to find the lowest soliton-kaon bound state we write the
kaon field as\cite{KM88,TSW94},
\be
\label{consatz}
K(\r,t) \ = \ k(r,t) \ \iso\cdot\piv_n(z) \ \chi \ ,
\ee
where $\chi$ is a two--component spinor.

     Using Eqs.(\ref{mansatz},\ref{consatz}) the explicit form of the
kaon-soliton effective lagrangian is
\be
\label{efflag}
L \ = \ \int \ dr \ r^2 \left \{ f_n(r) \ \dot{k\dag} \dot k - h_n(r) \
k^{\dagger\prime} k^\prime + i\lambda_n(r) (\dot{k\dag}k - k\dag \dot k) -
k\dag k(\mk2 + V^{eff}_n) \right \} \ ,
\ee
where
\be
f_n(r) \ = \ 1 + \fr{1}{e^2\fk2} \left({\df}^2 + 2 n\fr{\sin^2F}{r^2}
\right ) \ ,
\ee
and
\be
h_n(r) \ = \ 1 + \fr{2 n}{e^2\fk2} \fr{\sin^2F}{r^2} \ .
\ee
The term linear in time derivatives, whose coefficient is
\be
\lambda_n(r) \ = \ -\fr{n \ N_c}{2\pi^2\fk2} \
\df \ \fr{\sin^2F}{r^2} \ ,
\ee
is due to the Wess--Zumino action, and
\bea
 V^{eff}_n \ &=& \ \left [ \fr{2 n}{e^2\fk2 r^2} \left ( \cos^4F/2 -
2 \ \sin^2F \right ) - \fr{1}{4}\right ] \ {\df}^2 \nonumber \\
   &-&	   \fr{2}{r^2} \left ( \sin^2F - 4 \ \cos^4F/2 \right )
\left ( \fr{{\cal I}_n}{e^2\fk2}\fr{\sin^2F}{r^2} +
\fr{n}{4} \right )  \nonumber \\
   &-&	    \fr{6 n}{e^2\fk2 r^2} \fr{d}{dr}
\left [ \df \sin F \ \cos^2 F/2) \right ] -
\fr{1}{2}\fr{\fpi2}{\fk2}\mpi2(1 - \cos F) \ .
\eea
Here, the angular integrals ${\cal I}_n$ are defined as
\bea
{\cal I}_n&=& {1\over{4 \pi}} \int {2i\ dz d\bar z \over{ (1 + |z|^2)^2 }}
\ \left(
	 {{1 + |z|^2} \over{ 1 + |R_n|^2 }}
		  \left| {dR_n\over{dz}} \right| \right)^4 \ .
\eea

The diagonalization of the hamiltonian obtained from the effective
lagrangian Eq.(\ref{efflag}) leads to the kaon eigenvalue equation
\be
\left[ - {1\over{r^2}} \partial_r \left( r^2 h_n \partial_r \right)
+ m_K^2 + V^{eff}_n - f_n \epsilon_n^2	- 2 \ \lambda_n \ \epsilon_n
\right] k(r) = 0 \ .
\ee

In our numerical calculations we use two sets of parameters.
Set A corresponds to the case
of massless pions and Set B to the case where the pion mass takes its
empirical value $m_\pi = 138 \ MeV$. In both cases, $F_\pi$ and $e$
are adjusted so as to reproduce the empirical neutron and $\Delta$ masses.
The ratio $F_K/F_\pi$ and $m_K$ are taken at their empirical values,
$F_K/F_\pi = 1.22$ and $m_K = 495 \ MeV$. Once the parameters are
fixed we obtain the soliton profiles using for each baryon
number $B$ the rational map given in Ref.\cite{HMS97}. Then, we
proceed to solve the corresponding kaon eigenvalue equation.
Results are shown in Table 1. There we list the soliton mass (per
baryon number) and kaon eigenenergy for each baryon number for both
sets of parameters.

To get some insight on the quality of the ans\"atze we first concentrate
on $B=2$. In this case, the exact numerical
solution \cite{KS87} corresponds to an axially symmetric soliton where
the chiral angle $F$ depends not only on $r$ but also on $\theta$,
the angle with respect to the symmetry axis. A basic assumption of the
rational map ansatz is, however, that $F=F(r)$ for any winding number $B$.
In addition, within that particular class of solutions, this ansatz
proposes a very particular dependence of $\hat \pi_n$ as a function
of $\theta$ and $\phi$. For $B=2$, the best possible solution in which
the chiral angle depends only on $r$ was found in Ref.\cite{KKSO89} and
applied to strange dibaryons in Ref.\cite{TSW94}. Comparing with the results
of those references, we observe that
$M_{sol}$ as predicted by the rational map ansatz is around $20 \ MeV$
larger (for both sets of parameters) than the one in Ref.\cite{KKSO89}
\footnote{As compared to the exact numerically found minima, results in
Table 1 are around $50 \ MeV$ larger.}.
The effect on the kaon eigenenergy is even smaller:
the values of $\epsilon_2$ in Table 1 are only $6 \ MeV$
larger than those reported in Ref.\cite{TSW94}. Therefore,
possible improvements upon the rational map ans\"atze are not expected 
to change the values of the kaon eigenenergies in a significant way.

We turn now to the predictions for higher baryon numbers.
As a general trend we see that the kaon binding energies
$D^K_n = m_K - \epsilon_n$ decrease with increasing baryon number.
However, as in the case of energy required to liberate a single $B=1$ skyrmion
from the multisoliton background\cite{BTC90,BS97}, 
we observe some deviation from a smooth behaviour, 
namely, $D^K_4 > D^K_3$ and $D^K_7 > D^K_6$.
Consequently, such deviations will be also present in the multiskyrmion
mass per baryon. Interestingly, this kind of phenomena has been also observed
in some MIT bag model calculations\cite{FJ84}. There they are
due to shell effects.

Another magnitude of interest is the RMS baryonic radius of the
strange multiskyrmions with strangeness $S$. It is related to the isoscalar
electric radius and gives an idea of the size of these exotics.
It is given by
\be
\left( r^B_S \right)^2 = r^2_{sol} + S \ r^2_{K} \ ,
\ee
where $r^2_{sol}$ and $r^2_{K}$ are the solitonic and kaonic radii,
respectively, defined by
\bea
r^2_{sol} &=& - {2\over{\pi}} \int dr \ r^2 F' \sin^2 F  \ , \\
r^2_{K}   &=&  2 \int dr \ k^2 r^4 \left( f_n \epsilon_n +
\lambda_n \right) \ .
\eea
The corresponding results are given in Table 2 together with the radii
of the $S=-1$ multiskyrmions. As expected, both the solitonic and kaonic
radii increase monotonically as a function of $B$. For the radii of the
$S=-1$ multiskyrmions such increase is much slower as a result of a
compensation between both contributions. Actually, for Set A the radius
for $B=6$ is even smaller than that of $B=5$.
This effect, however, is very likely to be an artifact of this
parameter set since it does not appear for Set B.

To obtain states with definite spin and
isospin in the bound state model one should perform the $SU(2)$ collective 
quantization
of the soliton-meson bound system. Such quantization has to be done 
taking into account
all the symmetries of the bound system\cite{TSW94}. For $B>2$, however, this
is an intrincated problem which has not been fully understood even in the
case of non-strange multiskyrmions.
In this situation, we will only concentrate on estimating the centroid 
of the low-lying states assuming that, for those states, the rotational 
corrections are small. This approximation would become
better as $B$ increases due to the associated increase of the moments
of inertia.  Using this assumption we can address the issue of the
stability of the strange multibaryons against strong decays.
For the $H$-dibaryon we obtain
\be
M_H - 2 M_\Lambda = 12 \ MeV  \ .
\ee
Comparing with the results of Ref.\cite{TSW94} we note that
improvements on the description of the soliton background
together with the inclusion of rotational effects result in an
extra binding of $\approx 40 \ MeV$. In any case, given
the various uncertainties coming from i.e. Casimir effects, etc.,
it is clear that the present calculation basically leads
to the same qualitative result: within soliton models the $H$ is 
predicted to be very close to threshold. The situation 
concerning the stability of the tetralambda with respect
to the decay into two dilambda seems to be somewhat different,
however. From Table 1 we have
\be
M_{4\Lambda} - 2 M_{2\Lambda} = - 176  \ MeV \ ,
\ee
for both sets of parameters. Therefore, it is rather strongly bound.
It should be noticed that in this case improvements on the
ansatz will {\it decrease} the binding. For example, the solitonic
contribution in the present approximation is $-200 \ MeV$ to be
compared with the exact value $-150 \ MeV$ \cite{BS97}.
Still, the present results indicate that, within the Skyrme
model, the binding  of the tetralambda might be stronger
than previously expected\cite{IKS88}.

There are some other states with vanishing hypercharge which might be
stable against strong decays. To see that it is convenient to define
the ionization energy $I_B$ as
\be
I_{B} = M_{1} + M_{B-1} - M_B \ ,
\label{ion}
\ee
where $M_n$ is the mass of the state with $B=-S=n$. Within
the present approximations $M_n$ can be calculated from
Table 1 using $M_n = n ( M_{sol} + \epsilon_n) $. Results
for $B=5$ to 9 are shown in Table 3. We readily see that
$B=-S=5,8,9$ are unstable, $B=-S=6$ barely stable
while $B=-S=7$ is clearly stable against one hyperon
emission. It is not hard to check that the stability of
the $B=-S=7$ strange multiskyrmion is not violated by any
other strong decay. For example,
\be
M_{7\Lambda} - (M_{3\Lambda} + M_{4\Lambda}) = - 177  \ MeV \ .
\ee
Therefore, as it stands the present model predicts the existence of a
stable ``heptalambda".

In conclusion, we have studied strange multibaryon configurations
in the context of the bound state approach to the $SU(3)$ Skyrme
model using for the soliton background fields some ans\"atze based on
rational maps.
We have obtained the solutions of the kaon eigenvalue equation
for $B \le 9$ and $B=17$. With these solutions we have calculated
the associated baryonic radii. As expected, for $S = -1$ they
increase with increasing $B$.
We have calculated the strange multiskyrmion masses within the
adiabatic approximation. We have found that among the exotics with
vanishing hypercharge and low baryon number only those with $B=4$
and $7$ are clearly stable against strong decays. This might be of
interest for current experimental searches of strange exotics. 
Finally, we note that, in principle, it would be possible to study 
hypernuclei within
the present framework. However, as wellknown from the case of the
non-strange $B=2$ system, one should definitely perform a quantum
treatment of the $SU(2)$ multiskyrmion before a reasonable description
can be made. Investigations along this line are currently underway.

\vfill

\pagebreak

\begin{table}

\caption{
Soliton mass per baryon unit $M_{sol}$ and
kaon eigenenergy $\epsilon$ (both in $MeV$) as a
function of the baryon number $B$.
Set A corresponds to $F_\pi = 129 \ MeV$, $e= 5.45$, $m_\pi = 0$ while
Set B to $F_\pi = 108 \ MeV$, $e= 4.84$, $m_\pi = 138 \ MeV$. The other
parameters take their empirical values as it is explained in the text.}

\begin{center}
\begin{tabular}{|c|c|c|c|c|} \hline
 &  \multicolumn{2}{|c|}{Set A} &
    \multicolumn{2}{|c|}{Set B} \\ \hline
 $B$ & $M_{sol}$ & $\epsilon$ & $M_{sol}$ & $\epsilon$ \\ \hline
 $1$ & $863$ & $222$ & $864$ & $210$ \\
 $2$ & $847$ & $244$ & $848$ & $232$ \\
 $3$ & $830$ & $255$ & $832$ & $241$ \\
 $4$ & $797$ & $250$ & $798$ & $238$ \\
 $5$ & $804$ & $263$ & $808$ & $248$ \\
 $6$ & $797$ & $267$ & $802$ & $251$ \\
 $7$ & $776$ & $262$ & $780$ & $246$ \\
 $8$ & $784$ & $271$ & $790$ & $252$ \\
 $9$ & $787$ & $276$ & $796$ & $256$ \\
$17$ & $766$ & $284$ & $782$ & $256$ \\   \hline
\end{tabular}
\end{center}
\end{table}



\begin{table}
\caption{
Baryonic radii $r_B$ of the $S=-1$ multibaryons
as a function of the baryon number.
The solitonic $r_{sol}$ and kaonic $r_K$ contributions
to those radii are also displayed. Parameter sets A and B are defined
in the caption of Table 1. All the radii are in $fm$.
}

\begin{center}
\begin{tabular}{|c|c|c|c|c|c|c|} \hline
 &  \multicolumn{3}{|c|}{Set A} &
    \multicolumn{3}{|c|}{Set B} \\ \hline
 $B$ & $r_{sol}$ & $r_K$ & $r_B$ & $r_{sol}$ & $r_K$ & $r_B$ \\ \hline
 $1$ & $0.59$ & $0.44$ & $0.39$ & $0.68$ & $0.51$ & $0.44$ \\
 $2$ & $0.81$ & $0.65$ & $0.49$ & $0.94$ & $0.75$ & $0.57$ \\
 $3$ & $0.97$ & $0.80$ & $0.54$ & $1.12$ & $0.93$ & $0.63$ \\
 $4$ & $1.06$ & $0.89$ & $0.57$ & $1.23$ & $1.03$ & $0.67$ \\
 $5$ & $1.20$ & $1.03$ & $0.64$ & $1.37$ & $1.18$ & $0.68$ \\
 $6$ & $1.29$ & $1.13$ & $0.62$ & $1.48$ & $1.29$ & $0.73$ \\
 $7$ & $1.35$ & $1.18$ & $0.65$ & $1.55$ & $1.35$ & $0.76$ \\
 $8$ & $1.46$ & $1.29$ & $0.68$ & $1.66$ & $1.47$ & $0.78$ \\
 $9$ & $1.55$ & $1.39$ & $0.69$ & $1.76$ & $1.57$ & $0.80$ \\
$17$ & $2.05$ & $1.89$ & $0.79$ & $2.27$ & $2.09$ & $0.88$ \\	\hline
\end{tabular}
\end{center}
\end{table}

\begin{table}
\caption{Ionization energies $I_B$
as defined in Eq.(\ref{ion}) (in $MeV$) for $B=-S = 5$ up to $9$
multibaryons. Parameter sets A and B are as in Table 1.}

\begin{center}
\begin{tabular}{|c|c|c|} \hline
  B &  Set A & Set B \\ \hline
 $5$ & $- 62$ & $ -62  $  \\
 $6$ & $+ 36$ & $ +36  $  \\
 $7$ & $+ 203$ & $+ 210$  \\
 $8$ & $- 59$ &  $ - 80  $  \\
 $9$ & $- 42$ &  $- 58$  \\   \hline
\end{tabular}
\end{center}
\end{table}

\end{document}